# Determination of oxygen stoichiometry in the mixed-valent manganites

J.Yang [a], W. H. Song [a], Y. Q. Ma [a], R. L. Zhang [a], and Y. P. Sun [a, b,]*

[a] *Key Laboratory of Materials Physics, Institute of Solid State Physics, Chinese Academy of Sciences, Hefei, 230031, P. R. China*

[b] *National Laboratory of Solid State Microstructures, Nanjing University, Nanjing 210008, P. R. China*

## Abstract

The possible redox (oxidation reduction) chemical methods for precisely determining the oxygen content in the perovskite manganites including hole-doped $La_{1-x}Ca_xMnO_y$ and electron-doped $La_{1-x}Te_xMnO_y$ compounds are described. For manganites annealed at different temperatures, the oxygen content of the samples was determined by a redox back titration in which the powder samples taken in a quartz crucible were dissolved in (1+1) sulfuric acid containing an access of sodium oxalate, and the excess sodium oxalate was titrated with permanganate standard solution. The results indicate that the method is effective and highly reproducible. Moreover, the variation of oxygen content is also reflected in significantly affecting the electrical transport property of the samples, which is mainly considered to be closely related to introduce oxygen vacancies in the Mn-O-Mn network.

*Corresponding author. Phone: 86-551-5592757; Fax: 86-551-5591434; email: ypsun@issp.ac.cn

## 1. Introduction

Recently, the hole-doped manganese perovskites $Ln_{1-x}A_xMnO_3$ (Ln = La-Tb, and A = Ca, Sr, Ba, Pb, etc.) have attracted much renewed attention because of their importance for both fundamental issues in condensed-matter physics and the potential for applications [1-3]. Many studies indeed suggest that the ratio of $Mn^{3+}/Mn^{4+}$ is a key component for understanding the colossal magnetoresistance (CMR) effect and the transition from the ferromagnetic (FM) metal to paramagnetic (PM) semiconductor. Recently, many researches have placed emphasis on electron-doped compound such as $La_{1-x}Ce_xMnO_3$ [4-7] and $La_{1-x}Te_xMnO_3$ [8-10] because both electron as well as hole doped FM manganites may open up very interesting applications in the field of spintronics. It is well established that the electric transport and magnetic properties of manganites are closely related to the Mn oxidation state, which is determined by the oxygen content of the sample. Unfortunately, many of these 3d transition metal oxides possess a strong tendency for oxygen nonstoichiometry in the limit of the preparation method [11]. Therefore, it is crucially important to determine the exact oxygen content of the sample, not only for an accurate characterization of the nonstoichiometric oxide materials but also for a reasonable understanding of the physical properties as a result of the variation in the oxygen content. So far, a brief description of a method for determining the ratio of Mn (III) and Mn (IV) has been given in Ref.12,13, however, no quantitative data have been reported. Bloom et al. [14] described a quantitative method determining the amount of Mn (III) and Mn (IV). The solid sample is reacted directly with a reducing medium, ferrous ammonium sulfate. The equivalent of Fe(II) oxidized by Mn(III) and Mn(IV) are then determined by back titration with $KMnO_4$. In addition, Licci et al. mentioned a method determining manganese valence in complex La-Mn perovskites, which is based on two independent iodometric titrations, with amperometric dead-stop end-point detection [15]. The methods described above are quite complex. In this paper, we describe in detail a convenient and effective redox titration method for determining the oxygen content of perovskite CMR manganites including hole-doped and electron-doped manganites.

## 2. Experimental Details

### 2.1 Reagents and apparatus

Approximately 0.05N $KMnO_4$ solution was obtained by dissolving 0.8g AR-grade commercial salts in distilled water (500ml) and then boiled the solution for 1-2h under gently heating. The solution was deposited in shady place for a whole night and preserved them in dark containers after filtered impurity in the solution. About 0.075N $Na_2C_2O_4$ solution was obtained by dissolving dried salts (1g) in 1:1 v/v $H_2SO_4$ (200ml).

The concentration of $KMnO_4$ solution above made was anew decided before the solution was used to titrate. Approximately 20mL $Na_2C_2O_4$ standard solution was taken out using graduated flask and placed in 50 ml beaker. Then the solution was heated up 60-80 °C. At the same time, the $KMnO_4$ solution in the buret was dripped drop by drop into the $Na_2C_2O_4$ solution containing $H_2SO_4$ until a pink color in the $KMnO_4$ solution can keep 30 seconds and the temperature of the solution for the end-point of titration is not under 60 °C. Thus the concentration of $KMnO_4$ solution can be calculated as 0.086N according to the dosage of the $Na_2C_2O_4$ standard solution.

The $La_{0.7}Ca_{0.3}MnO_y$ and $La_{0.9}Te_{0.1}MnO_y$ reference samples were synthesized through solid-state reaction from a stoichiometric mixture of high-purity powders of $La_2O_3$, $CaCO_3$, $MnO_2$ and $La_2O_3$, $TeO_2$, $MnO_2$, respectively. The powders obtained were ground, palletized, and sintered at 1400 °C and 1030 °C, respectively, for 24h with several intermediate grindings. In order to vary the oxygen content of the samples, we annealed the samples at 750 °C, 800 °C and 850 °C, respectively, in $N_2$ atmosphere under 2MPa pressure for 4h with graphite powder placed near the samples. The powder x-ray diffraction at room temperature shows that all the samples are single phase with no detectable secondary phases.

## 2.2 Analytical procedure

Approximately 0.2 g of the powder samples were weighed and placed in a quartz crucible, then dissolved in (1+1) sulfuric acid containing an access of sodium oxalate (20ml). Note that the crucible containing the power samples was sealed through lidding on and placed in a bigger crucible, which was also covered with a crucible lid. The sealed crucible was laid in an oven under the temperature of 50 °C for 10h. So the powder samples for the hole-doped manganites $La_{0.7}Ca_{0.3}MnO_y$ reacted with a suitable excess of $Na_2C_2O_4$ (~0.1g for 0.2g of sample), which deoxidize $Mn^{3+}$ and $Mn^{4+}$ to $Mn^{2+}$ according to the quantitative reaction:

$$Mn^{3+} + C_2O_4^{2-} \rightarrow Mn^{2+} + 2CO + O_2, \qquad (1)$$

$$2Mn^{4+} + C_2O_4^{2-} \rightarrow 2Mn^{2+} + 2CO + O_2 \qquad (2)$$

For the electron-doped manganites $La_{0.9}Te_{0.1}MnO_y$, the reaction (2) does not occur due to the mixed-valence state of $Mn^{3+}$ and $Mn^{2+}$.

When the time of heat preservation was over, the crucible was taken out and the solution in it was moved into a clean and dried beaker (400ml), and then added the distilled water till 150ml. The beaker was placed on the magnetic heating stirrer and stirred for about 10 minutes. Then the $KMnO_4$ solution above prepared in the buret was dripped drop by drop into the beaker until a pink color in the solution could keep 30 seconds and the temperature of the solution for the end-point of titration was not under 60 °C. Note that the titration was carried out in the process of heating and stirring at one time. The unreacted $Na_2C_2O_4$ after finishing reaction (1) and (2) oxidized $Mn^{7+}$ to $Mn^{2+}$ according to the quantitative reaction:

$$5C_2O_4^{2-} + 2MnO_4^{-} + 16H^{+} \rightarrow 2Mn^{2+} + 10CO_2 + 8H_2O. \quad (3)$$

So the oxygen content for the samples $La_{1-x}Ca_xMnO_y$ is then obtained as

$$y = \frac{5-x}{2} + \frac{n_1 - n_2}{n_3}, \quad (4)$$

and the oxygen content for the samples $La_{1-x}Te_xMnO_y$ can be calculated as follows:

$$y = \frac{5+x}{2} + \frac{n_1 - n_2}{n_3}, \quad (5)$$

where $n_1$ is the amount of total $Na_2C_2O_4$, $n_2$ is the amount of $Na_2C_2O_4$ reacted with $KMnO_4$ and $n_3$ is the amount of the powder samples.

## 3. Results and discussion

The reliability of the procedure for determining the oxygen content of the sample was examined by analyzing comparison samples of reagent grade $MnO_2$ and $KMnO_4$. The results are shown in Table 1.When the measurement is repeated for three times on the same material, reproducibility of results is better than 1%. As one can see, the reagent grade $MnO_2$ contained 4.21% Mn (III) and the determination of Mn (VII) in $KMnO_4$ is in good agreement with theory. The oxygen content and average Mn valence of the samples $La_{0.7}Ca_{0.3}MnO_y$ and $La_{0.9}Te_{0.1}MnO_y$ under different annealing temperatures is measured by the method described above and the results are summarized in Table 2. As we can see, the oxygen stoichiometry decreases with the increase of the annealed temperature, which is consistent with the results reported in Ref. 16,17. Besides, we find the temperature of end-point for titration is a main factor affecting the oxygen content of the samples. The results show that the oxygen content of the samples systematically decreases with increasing the temperature of end-point for titration and the maximum difference of the oxygen content according to two extreme temperatures, i.e., 60 °C and 80 °C, is below 1%. Here we adopt the average value for the results. Other errors are negligible.

In order to confirm the validity of the analytical procedure, the resistance of the samples $La_{0.9}Te_{0.1}MnO_y$ as a function of temperature was measured by the standard four-probe method from 25 to 300K. Fig.1 shows the temperature dependence of resistivity for the sample of as-prepared $La_{0.9}Te_{0.1}MnO_y$ (curve A), the sample annealed at 750 °C (curve B), annealed at 800 °C (curve C), and annealed at 850 °C (curve D) for 4h in $N_2$ with graphite powders nearby, respectively. For sample A, it shows that there exists an insulator-metal (I-M) transition at $T_{P1}$ (= 246 K). In addition, there exists a bump shoulder at $T_{P2}$ (= 223 K) below $T_{P1}$, which is similar to the double peak behavior observed usually in alkaline-earth-metal-doped and alkali-metal-doped samples of $LaMnO_3$ [18-22]. As to the origin of double peak behavior in ρ(T) observed in hole-doped manganites, several models including the spin-dependent interfacial tunneling due to the difference in magnetic order between surface and core, [20] which is intimately related to the size of grains, magnetic inhomogeneity [21] etc. have been proposed to interpret it. However, its real origin is not very clear at present. More interesting phenomenon is that double I-M transitions occur significant variation with the annealed temperature. For the sample annealed at 750 °C (sample B), the double I-M transitions shift to low temperatures, which has $T_{P1}$= 240 K and $T_{P2}$= 205K. Additionally, I-M transition at $T_{P1}$ becomes weak and I-M transition at $T_{P2}$ becomes more obvious behaving as the noticeable character of peak compared with the character of bump shoulder of the as-prepared sample. For the sample C and D, ρ(T) curves display the semiconducting behavior ( $d\rho/dT < 0$ ) in the whole measurement temperature range and the resistivity maximum is of about 6 orders of magnitude larger than that of the as-prepared sample. It should be mentioned that for $La_{0.9}Te_{0.1}MnO_{2.97}$ (sample B), the ratio of $Mn^{2+}/(Mn^{2+}+Mn^{3+})$ is close to 16%, comparable to that of $La_{0.84}Te_{0.16}MnO_3$. From the known experimental data [8], such a system should show higher I-M transition temperature and lower resistivity. For $La_{0.9}Te_{0.1}MnO_{2.86}$ (sample C) and $La_{0.9}Te_{0.1}MnO_{2.83}$ (sample D), the ratio of $Mn^{2+}/(Mn^{2+}+Mn^{3+})$ is close to 38% and 44%, comparable to that of $La_{0.62}Te_{0.38}MnO_3$ and $La_{0.56}Te_{0.44}MnO_3$, respectively. In our previous work [23], such two systems should show an insulator-metal transition. However, the case is clearly not observed in sample C and D. So oxygen content reduction in $La_{0.9}Te_{0.1}MnO_y$ is expected to cause two effects. One is the increase in the $Mn^{2+}/Mn^{3+}$ ratio, driving the carrier density increase, and causing the decrease of the resistivity. Another effect is the occurrence of the local lattice distortion due to the introduction of oxygen vacancies in the Mn-O-Mn network, which is important for electrical conduction. The local lattice distortion caused by oxygen vacancies in samples is confirmed by the structural parameter fitting through the Reitveld technique. The local lattice distortion is also expected to increase the resistivity due to the reduced $e_g$ electron bandwidth [16-17]. As a result, the competition of the two effects suggested above decide the behavior of resistivity and the second factor get the upper hand of the first one. Therefore, the increase of resistivity of the samples can be attributed to the reduction of oxygen content in $La_{0.9}Te_{0.1}MnO_y$. In addition, the variation of oxygen content of the samples is also reflected in the effect on the magnetic properties of the samples. (not shown here).

In conclusion, we develop a reliable and effective redox chemical method, which can be used to determine conveniently, and effectively the oxygen-content of manganites including the hole-doped and electron-doped CMR materials. The variation of oxygen content is also reflected in the effect on the electrical transport and magnetic properties of the samples

**ACKNOWLEDGMENTS**

This work was supported by the National Key Research under contract No.001CB610604, and the National Nature Science Foundation of China under contract No.10174085, Anhui Province NSF Grant No.03046201 and the Fundamental Bureau Chinese Academy of Sciences.

**Table 1. Analysis of known compounds**

| Compound | Mn (III) (%) | Mn (IV) (%) | Mn (VII) (%) | Mn (%) (theoretical) |
|---|---|---|---|---|
| $MnO_2$ | 4.21 | 58.98 | – | 63.19, (IV) |
| $KMnO_4$ | – | – | 34.51 | 34.77, (VII) |

**Table 2. The oxygen content in the synthetic samples**

| Sample | Material | Treatment temperature (°C) | Calculated oxygen content | Average Mn valence |
|---|---|---|---|---|
| 1 | $La_{0.7}Ca_{0.3}MnO_y$ | As-prepared | 3.02 | 3.34 |
| 2 | $La_{0.7}Ca_{0.3}MnO_y$ | 750 | 2.98 | 3.26 |
| 3 | $La_{0.7}Ca_{0.3}MnO_y$ | 800 | 2.88 | 3.06 |
| 4 | $La_{0.7}Ca_{0.3}MnO_y$ | 850 | 2.85 | 3.00 |
| 5 | $La_{0.9}Te_{0.1}MnO_y$ | As-prepared | 3.01 | 2.92 |
| 6 | $La_{0.9}Te_{0.1}MnO_y$ | 750 | 2.97 | 2.84 |
| 7 | $La_{0.9}Te_{0.1}MnO_y$ | 800 | 2.86 | 2.62 |
| 8 | $La_{0.9}Te_{0.1}MnO_y$ | 850 | 2.83 | 2.56 |

## Figure captions

Fig.1 Temperature dependence of resistivity for the as-prepared (sample A), 750 °C annealed (sample B), 800 °C annealed (sample C), and 850 °C (sample D) samples, respectively.

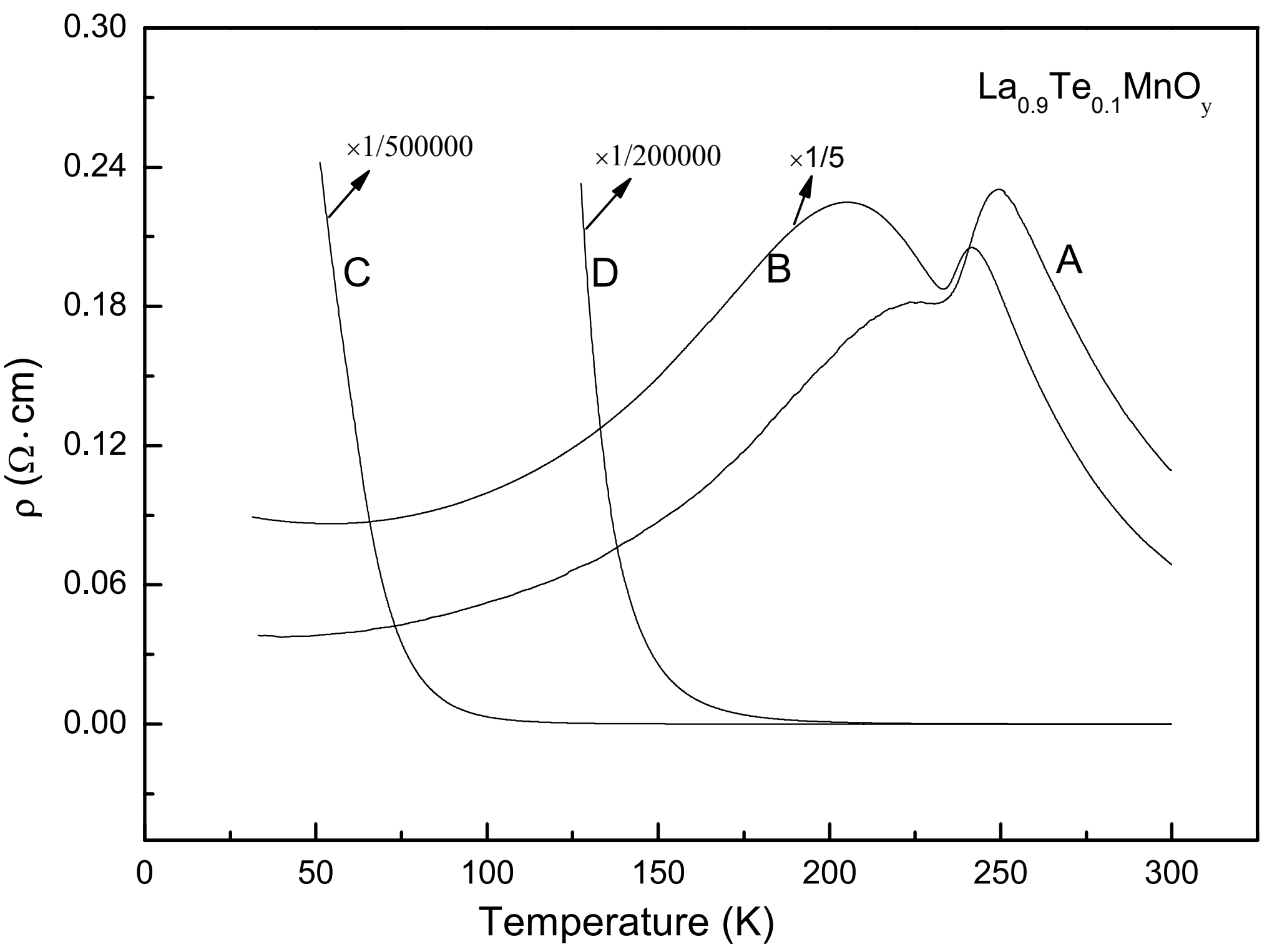

Fig.1 J. Yang et al.,